\documentclass[
notitlepage, 
onecolumn, 
pre,
aps,
superscriptaddress,
preprintnumbers,
amsmath,
amssymb]{revtex4-1}

\usepackage{amsthm,amsmath}
\RequirePackage{hyperref}
\usepackage[utf8]{inputenc} 

\usepackage{graphicx, subfigure,microtype}
\usepackage{amssymb, amsmath}
\usepackage{amsfonts,bbm}
\usepackage{mathrsfs}
\usepackage{fourier}
\usepackage{verbatim}

\usepackage{url}
\usepackage{hyperref}

\usepackage{tikz}
\usetikzlibrary{positioning}
\usetikzlibrary{plotmarks}

\usepackage{color}

\usepackage{amsthm}

\theoremstyle{plain} 

\theoremstyle{definition}

\theoremstyle{remark}
\newtheorem{rmk}{Remark}

\begin{document}
\title{Classes of random walks on temporal networks with competing timescales}

\author{Julien Petit}
\email{julien.petit@unamur.be}
\affiliation{Department of Mathematics, Royal Military Academy, Brussels (Belgium)}
\affiliation{naXys, Namur Institute for Complex Systems, Namur (Belgium)}%
 
 \author{Renaud Lambiotte}%
 \affiliation{Mathematical Institute, University of Oxford, Oxford (UK)}%

\author{Timoteo Carletti}
\affiliation{naXys, Namur Institute for Complex Systems, Namur (Belgium)}

\date{\today}

\begin{abstract} 
Random walks  find  applications in many areas of science and are the heart of essential network analytic tools. When defined on temporal networks, even basic random walk models may exhibit a rich spectrum of behaviours, due to the co-existence of different timescales in the system. Here, we introduce random walks on general stochastic temporal networks allowing for lasting interactions, with up to three competing timescales. We then compare the mean resting time and stationary state of different  models. We  also discuss the accuracy of the mathematical analysis depending on the random walk model and the structure of the underlying network, and pay particular attention to the emergence of non-Markovian behaviour, even when all dynamical entities are governed by memoryless distributions. \\[1.5em]
\end{abstract}

\maketitle

\section{Introduction}

Diffusion on static networks is a well-known, extensively studied problem. Its archetype is the study of random walk processes on networks, that are essentially equivalent to Markov chains, and are used as a baseline model for the diffusion of items or ideas on networks \cite{masuda2017random}, but also as a tool to characterise certain aspects of their organisation such as the centrality of nodes \cite{brin1998anatomy,langville2004deeper,lambiotte2012ranking} or the presence of communities \cite{rosvall2009map,delvenne2010stability}. Random walks find applications in biology, particle physics, financial markets, and many other fields. This diversity contributes to the many existing variants of random walks, including Levy flights~\cite{klafter2011first}, correlated walks~\cite{renshaw1981correlated}, elephant walks~\cite{schutz2004elephants},  random walks in heterogeneous media~\cite{grebenkov2018heterogeneous}, in crowded conditions~\cite{asllani2018hopping}, or even quantum walks~\cite{kempe2003quantum} (see \cite{klafter2011first,hughes1995random} and the many references therein).  In real world scenarios, a core assumption of random walk models, i.e. that the network is a static entity,  is often violated~\cite{holme2015modern,masuda2016guidance}. Empirical evidence shows instead that the network should be regarded as a dynamical entity, and part of the research on random walks is devoted to their dynamics on temporal networks ~\cite{perra2012random,starnini2012random,petit2018random}. The  distinction between the dynamics on the network and the dynamics of the network is not always  clear, which naturally lead to a standard classification of the different types of random-walk processes~\cite{masuda2017random}. 
According to this classification, there are two dominant facets of random walks on temporal networks. 

Firstly, it is relevant for many dynamical systems on networks, including random walks,   to distinguish between node-based and edge-based dynamics~\cite{Porter_2016}. In a node-centric walk, a stochastic process occurring at the level of the node  determines  the duration before the next jump. At that moment,  the walker can move to a new destination, selecting any of the outgoing edges. At variance, in an edge-centric model, the links are the driving units. In other words, the links become available for transport, then vanish, according to their stochastic process. The edge-centric model is also known under the intuitive denomination of fluid model.

Secondly, one can draw a line between active and passive models. The walk is called active when the waiting time before the next jump is reset by each jump of the walker. Active walks are common models for animal trajectories. On the other hand, in passive models the motion of the walker is instead constrained by the temporal patterns of (typically the edges of) the network. An example would be that of a person randomly exploring a public transportation network, taking every available ride.

In short, there are clocks, either on the nodes or on the edges (node-centric vs edge-centric models), and either the clocks are reset following each jump of the walker or they evolve in an independent manner (active vs passive models). However, this classification is too restrictive when it comes to scenarios where the walker has an own dynamics, which is decoupled from the transport layer represented by the network. More precisely, a first objective of this work is to formulate natural extensions of existing random walks, that   arise when the network is such that the duration of the contacts between the nodes is not instantaneous~\cite{gauvin2013activity,sekara2016fundamental}. This is in contrast with a majority of approaches which assume that the network evolution can be modelled as a point process~\cite{hoffmann2012generalized}. On top of that, the walker does not necessarily jump through an available link. Instead, it is constrained by its own waiting time after each jump. By doing so, we allow for the possibility to investigate the importance of a timescale associated to edge duration, which may appear in certain empirical scenarios, e.g. for phone calls versus text message communication. This importance is assessed by making a comparison with the timescale of the walker's self-imposed waiting time. We will show that the diffusive process can effectively be  a combination of active and passive diffusion, with clocks both on the edges (determining the network's contribution to the overall dynamics) and on the nodes (determining the walker's contribution). 
As a next step,  we study in detail the mean residence time for the different models, with the motivation that  this is an appropriate measure for the speed of spreading processes.  Because random walks can be used to determine the most central nodes in a network, a final application will be the analysis of the rankings of the nodes resulting from the different models.  

Overall, our main message is that diffusion on temporal networks is a question of timescales and that simplified existing models may emerge and provide accurate predictions in certain regimes when some processes can be neglected. But we also show that in other regimes, the  models that go beyond the usual binary classification (active or passive, node-centric or edge-centric) are relevant. 
The interaction between dynamical parameters (which determine the waiting time densities for the walker and the network) and topology parameters (such as vertex degrees) induces non trivial effects on - for instance - the steady state.
Our results also show that the temporality of the underlying network may lead to a loss of Markovianity for the dynamics of the walker. This non-Markovianity may reveal itself in  the properties of the timings at which events take place or in trajectories that can thus only be approximated by first-order Markov processes.

Our approach is based on the generalized integro-differential master  equation obtained for non-Poisson random walks on temporal networks~\cite{hoffmann2012generalized}. Using this equation, it becomes apparent that we  need to determine the distribution of the total waiting time (or residence time) on the node. By direct integration, we then obtain the mean residence time. We can also compare the different models, and evaluate the dominating timescales in  regimes of extreme values for the dynamical parameters. Finally, combining an asymptotic analysis of the master equation and the residence times leads to the steady state for the different models. The analysis up to that point does not take into account a memory effect arising from walker-network interaction. As stated above, this memory effect means that the trajectories followed by the walker cannot be described by a Markov chain because of correlations between the timings or directions of the jumps. This will be discussed based on the approach of~\cite{petit2018random}.

The structure of the paper reflects this approach. In section~\ref{sec:models} we introduce the models with lasting edges that complement the classical active or passive, node-centric or edge-centric models which all have instantaneous contacts. The master equation from which the analysis is drawn follows in section~\ref{sec:masterequation}. The mean residence time of the models and an analysis of competing timescales is the subject of section~\ref{sec:residence times}, followed by the steady states in section~\ref{sec:steady-state}. Section~\ref{sec:memory} addresses the memory effect due to walker-network interaction before we conclude in section~\ref{sec:conclusion}.

\section{Classification of models with two or three timescales}
\label{sec:models}
\begin{figure}[h!]
	\centering
	\includegraphics[width=.95\textwidth]{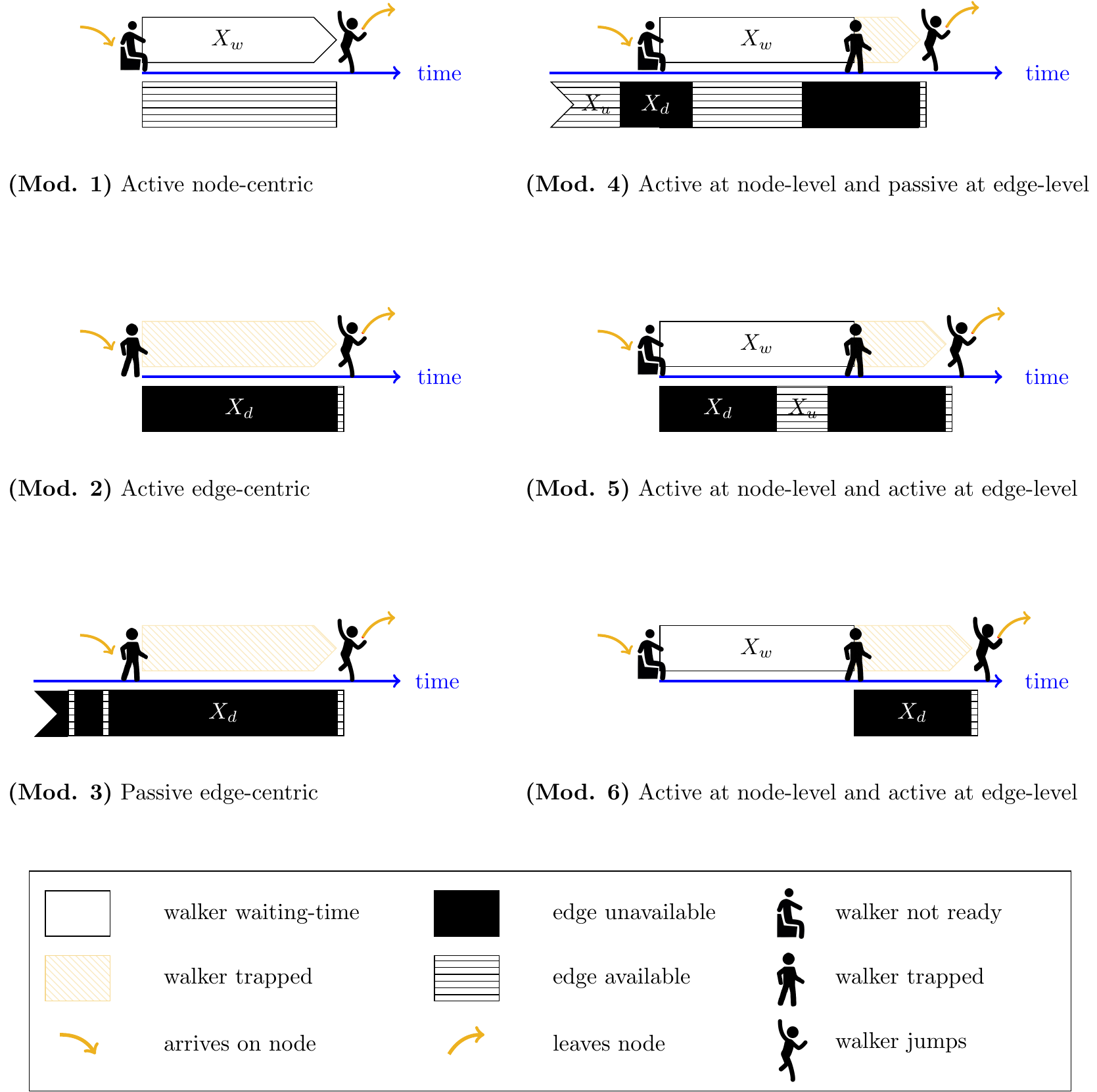}
	\caption{\textbf{Random walk models.}  Each of   the six panels represents a model of random walk.  In each case, the ribbon above the blue time axis represents the dynamics at node level, that is, the waiting time $X_w$ of the walker, possibly extended by a period where the walker is trapped. The ribbon under the time axis represents the dynamics at edge-level, with the associated random variables $X_u$ and $X_d$ for the up-times and up-times respectively. In the left column, the three classical models with only one timescale are represented, whereas in the right column, the walks have two (model 6) or three  (models 4 and 5) competing timescales. For each model, the arrival time on the node is located on the time axis by the leftmost man icon (standing or sitting), and the time of the jump corresponds to the rightmost jumping-man icon.   When the node-level  ribbon (above the blue axis) and the edge-level ribbon (below the blue axis) have a synchronous start from the left, as in models 1, 2 and 5,  the walk is fully active. Model  6 is also fully active, even though the reset of the  edges happens after $X_w$.  On the other hand,  models 3 and 4 are passive at edge-level. This means that the edges dynamics is independent of the walker, in the sense that the arrival of the walker on  a node does not mean the state of outgoing edges is reset.  }
	\label{fig:models}
\end{figure}

The goal of this section is  to present and to classify random walk models with up to three timescales : one associated to the walker, and two with the network. Let us first briefly review the three   classical models of continuous-time random walks. Their respective sets of microscopic motion rules are described by the  three panels in the left column of figure~\ref{fig:models}. 
\begin{itemize}
	\item The first walk,  labelled (Mod. 1), is   an active node-centric model where the walker resets the clock of a node upon arrival on the node.  One can think of the clock as being attached to the walker, and as obeying the walker's dynamics, whereas the network is then  static. The waiting time on a node corresponds to the random variable $X_w$. In the simplest case, $X_w$ follows an exponential distribution with probability density function (PDF) $\psi(t) = \mu e^{-\mu t}$, and the sequence of samples of $X_w$ follows a Poisson process. In general cases however, the distribution for $X_w$  is not exponential and the sequence of durations is a renewal process. In this work, we will concentrate on the simpler, exponential case allowing us to bring out effects due to walker-network interaction.  Summing up, the variable $X_w$ represents the active, node-level feature in all models. 
	
	\item The second walk with label (Mod. 2), is the active edge-centric model. The walker accordingly resets the down-time of the edges leaving a node, upon arrival on that node. This down-time is the random variable $X_d$ and  determines the period of unavailability of the edge.
	\item The third walk,  (Mod. 3), is again an edge-centric model with a passive walker, who passively follows edge activations.
	\item  The fourth combination corresponding to the two classification criteria,  the passive node-centric walk, appears to be irrelevant for practical applications~\cite{masuda2017random}. We won't discuss it further. 
\end{itemize}

Those three models  possess a single timescale, associated to the clock of the walker or of the edges. 
We now introduce three natural extensions, each featuring an active walker, carrying its own clock. This clock is reset after each jump, and determines a walker's self-imposed waiting time. We consider different behaviours for the transport layer, which will define the edge-level dynamics of the models. 
\begin{itemize}
	\item In a first declination called model 4, the network functions independently of the walker. Hence, the edge-level dynamics is passive.  In this model, each edge cycles through states of availability followed by periods of unavailability. The durations of the up-times is a random variable $X_u$, and the durations of the down-times is the already-introduced random variable $X_d$. Again, we assume both are exponentially distributed, with respective PDF's $U(t) = \eta e^{-\eta t}$ and $D(t) = \lambda e^{-\lambda t}$.   As depicted by figure~\ref{fig:models}, when the walker is ready to jump at the end of $X_w$, there are two possibilities. If one or more outgoing edges are available, a new destination node is selected randomly without bias. The total waiting time on the node is thus $X_w$. If no outgoing edge is available, then the walker is   trapped on the node, and will wait until the next activation of an edge.
	Observe that a similar\footnote{%
		Another behaviour would have been that the ready-to-jump but trapped walker  waits for another period drawn again from the  distribution of $X_w$,  before attempting another jump. In this scenario, the induced delay before the jump also depends  on the dynamics of the walker and not only on that of the  network  through availability of edges at the end of the prolonged stay~\cite{figueiredo2012characterizing}. With this choice, the analysis of the model would follow the same steps. } behaviour for the trapped walker was assumed and some consequences analysed in~\cite{petit2018random}.  This model has three timescales\footnote{We implicitly assume that the timescales are well-defined and are correctly represented by the expectation of the random variables. This assumption holds for exponential distributions, but wouldn't apply for power-law distributions. }, one for the walker ($X_w$) and two for the network ($X_u$ and~$X_d$). 
	
	\item In a second declination called model 5, the walker this time has an active role with respect to the network. At the beginning of the walker's waiting time $X_w$, the state of each outoing edge is reset to unavailable, for a duration $X_d$. Then follows a period of availability $X_u$, then a down-time, and so on, until the walker is ready and an edge is available. This model is active at node- and edge-levels, and also has three timescales.
	
	\item The third declination, model 6,  is similar to model 5 in the sense that the walker also actively resets the transport layer, but this reset occurs at the end of the walker's waiting time $X_w$. Therefore, the walker is always trapped for a duration $X_d$ and the dynamics only has two timescales, one for the walker and one for the down-times.	
\end{itemize} 

\begin{figure}
	\centering
	\includegraphics[width=.8\textwidth]{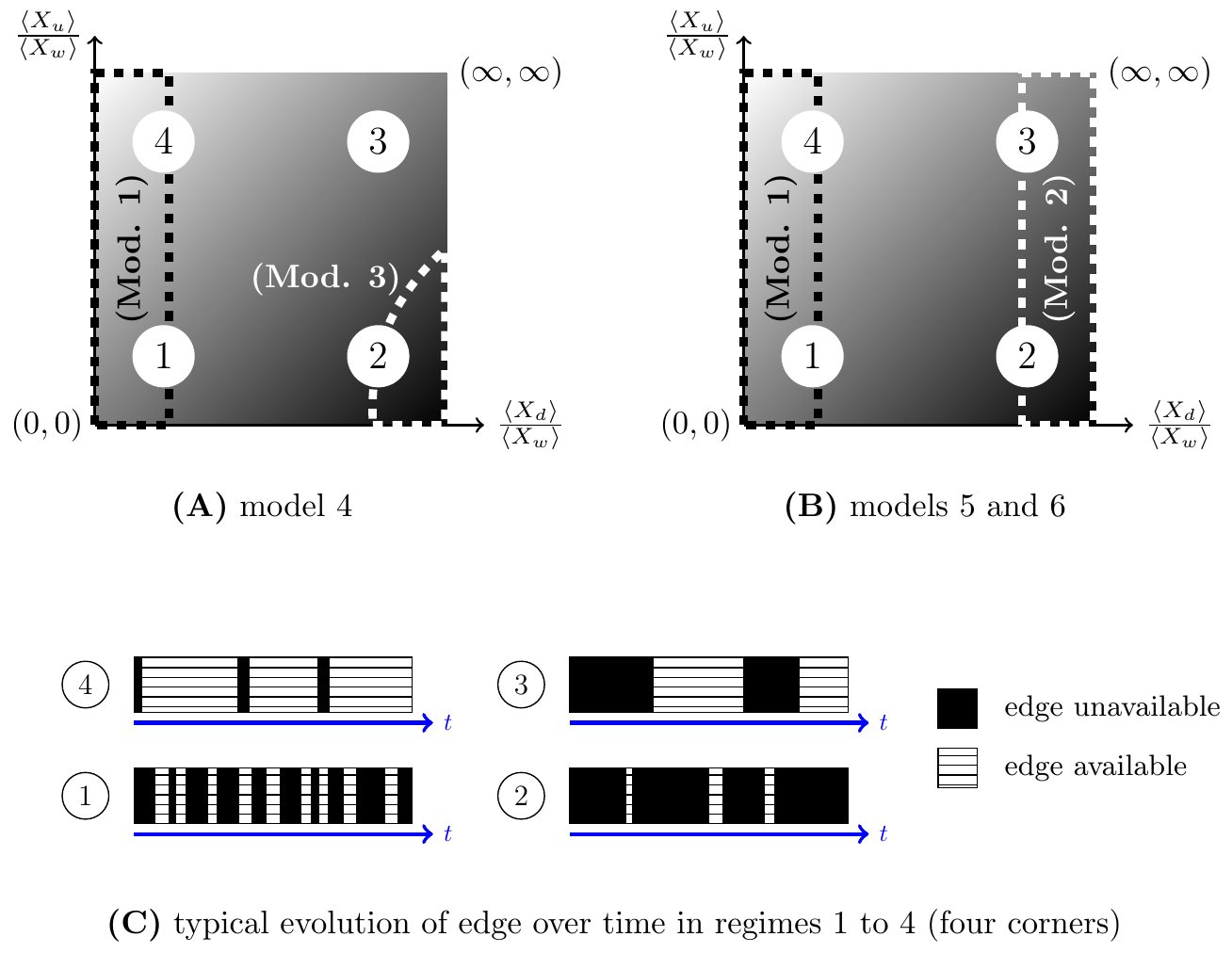}
	\caption{\textbf{Choice of modelling according to the competing timescales.}   When the timescale of each of the dynamics corresponding to $X_w$, $X_u$ and $X_d$ is well-defined, as is the case with exponential distributions,  models 4, 5 and 6 may be approached by the simpler models 1, 2 and 3. The regions where those approximations are valid are indicated on panel (A) for model 4, and on panel (B) for models 5 and 6. This qualitative idea of equivalence between models is revisited more rigorously in figure~\ref{fig:arrows}.}
	\label{fig:cartoon}
\end{figure}

The set of motion rules for three versions of walks on networks with duration of contacts have now been presented. The mathematical modelling is expectedly more involved than in the single-timescale models 1, 2 or 3.  However, it is not needed to always consider the models in their full complexity. Indeed, it results from a close look at  figure~\ref{fig:models} that in limiting regimes,  the models approach the classical node- or edge-centric walks. This   qualitative observation is presented on figure~\ref{fig:cartoon}. The  analysis of  section~\ref{sec:residence times} will provide a quantitative justification of this figure.

We can proceed with the analysis based on the master equations in the next section.

\section{Master equation}\label{sec:masterequation}
Starting from  the microscopic mobility rules defining the model, a master equation for the evolution of the walker's position across the network can be derived. This position is encoded in the row vector
\begin{equation*}
n(t) = \left(  n_1(t), n_2(t), \ldots, n_N(t)    \right), 
\end{equation*}
describing the residence probabilities on the nodes. Here, $N$ is the number of nodes of the network and   each component $n_i(t)$ gives the probability that the walker is located on node $i$ at time $t$.  Once the equation is obtained, the analysis of various properties of the walk becomes accessible.

The master equation differs based on whether there is memory in the process or not, justifying the following distinction. 

\subsection{Markovian vs non-Markovian walks}  

If the process is   Markovian, the trajectory starting from a given time only depends on the state of the system at that time, and the equation simplifies, typically under the form of a differential equation. In that case, the distribution of the time that a walker still remains on a node  is not influenced by the time spent so far on the node. Otherwise, in the non-Markovian case, the master equation has a more complex but still useful form.  Two different types of non-Markovianity may emerge in the system: a non-Markovianity  in time, when memory affects the timings of future jumps, or in trajectory, when   memory impacts the choice of the next destination node. Non-Markovianity in time means that the number of jumps of the process deviates from a Poisson process. For instance, this happens in a node-centric walk where the waiting time before activation is non-exponential. Non-Markovianity in trajectory means that the paths followed by the walker cannot be described by a Markov chain because of  correlations between edge activation along this trajectory, even if edges are statistically  independent.

\subsection{Markovian random walks}
The active node-centric walk (Mod. 1) is arguably the simplest to study among continuous-time random walks. It is Markovian  when the waiting time on the node is exponentially distributed. To fix notations, consider  a  directed, strongly connected\footnote{This means that there exists a path along connected edges allowing to reach any node  of the graph from any other node~\cite{newman2018networks}. } graph  $\mathcal G = (V,E)$ on $N$ nodes, where  $E$ is the set of allowed directed edges between pairs of nodes of the set $V$.  Let $V_j$ be the set of nodes reachable from node $j$ in $\mathcal{G}$. 
When the   outgoing edges are chosen uniformly by the walker, the walk on $\mathcal G$ is governed by the well-known master  equation for the residence probabilities~\cite{angstmann2013pattern}: 
\begin{equation} \label{eq:angstmann}
\dot n_i(t) = \sum_{j = 1}^{N} \mu_j n_j \frac{1}{k_j} A_{ji} - \mu_i n_i, 
\end{equation}
where $A_{ij}$ is the adjacency matrix ($A_{ij }= 1$ if  $i\rightarrow j$ is an edge of $E$, and is zero otherwise),  $k_j = |V_j|$ is the cardinal of $V_j$, namely, the positive out-degree of node~$j$, and where $\mu_j$ is the exponential rate in the node. 
If the rate is the same on all nodes, $\mu_j = \mu$ for all $j$, and after scaling of the time variable $t \mapsto \mu t$, equation~\eqref{eq:angstmann} reads 
\begin{equation}
\label{eq:node-centric}
\dot n = n D^{-1} (A-D) = nL^{rw}, 
\end{equation}
where $D$ denotes the diagonal matrix of the out-degrees. The matrix $L^{rw} = D^{-1}A -I$ is the so-called random walk Laplacian. 

Observe that an alternative modelling for the process of  model~1 is to consider that the walker is always ready to jump, and that the edges activate like in the edge centric walk~(Mod. 2). To illustrate this fact, let us start from model 1 where we assume that the rates on the nodes are proportional to degree : $\mu_j = \lambda k_j$,  so that the rate of jump across each edge of the graph is the same, independently of the degrees of the nodes. The residence time on  node $j$ is   the random variable $X^{(j)}$ which is here written as
\begin{equation}
\label{eq:min}
\mathbb X^{(j)}_d= \min \left\{  \left. X_d^{\ell \leftarrow j} \, \right \vert \ell \in V_j   \right\}  , 
\end{equation}
namely the minimum of $k_j$ exponential distributions with rate $\lambda$, which again follows an exponential distribution with rate $k_j \lambda$.  Equation~\eqref{eq:angstmann} becomes
\begin{equation} \label{eq:angstmann2}
\dot n_i(t) =  \sum_{j = 1}^N n_j\lambda A_{ji} - \lambda k_i   n_i. 
\end{equation}
In matrix form, recalling that $n$ is a row vector we have $\dot n = \lambda n(A-D)$, or again, after scaling of time with respect to $1/\lambda$, 
\begin{equation}
\label{eq:edge-centric}
\dot n =  n L, 
\end{equation} 
where this time $L = A-D$ is known as the graph or combinatorial Laplacian. Equation~\ref{eq:edge-centric} is well known to correspond to model 2, where $\lambda$ would be the rate of the exponential distribution governing the time before activation of each edge.  This shows that the same process  - in terms of trajectories - can be seen as happening atop of a static graph,  or a temporal graph. The walk on a static graph generates a temporal network where jumps across edges are considered as edges activation. Note again that this freedom of the choice of modelling  exists thanks to the restrictive framework of exponential distributions.

\begin{rmk}\label{rmk:switch}(On random walk time-dependent Laplacian)
	In previous works dealing with  synchronisation~\cite{stilwell2006sufficient,belykh2004blinking}, desynchronisation~\cite{lucas2018desynchronization}, instabilities in reaction-diffusion systems~\cite{petit2017theory} and other works about dynamical systems on time-varying networks~\cite{zanette2004dynamical,holme2015modern}, a time-dependent Laplacian $L(t)$ has replaced the usual graph Laplacian $L$ in the equations.  Here we want to comment on \eqref{eq:edge-centric} in this time-varying setting, that is, 
	\begin{equation}\label{eq:temporal-laplacian}
	\dot n = n L(t). 
	\end{equation}
	In our framework, we can consider this equation as associated to a passive edge-centric walk on  switched networks, where the underlying network of possible links varies in time. The rewiring occurs for several edges simultaneously at discrete time steps, as opposed to the continuous-time process we have considered so far, where no two edges can change states at the same time.  The adjacency matrix is $A(t) = A^{\left[\chi(t)\right]}$, where $\chi : \mathbb{ R}^+ \rightarrow I \subset \mathbb{ N}$ selects one possible   graph configurations in the set $ \left\{    A^{\left[ i \right]}   \right\}_{i \in I}$. The definition of the Laplacian is then $L(t)  = A(t) - D(t)$, where $D(t)$ contains the time-dependent degrees on its diagonal. Note that for simplicity, we have again assumed that the rate $\lambda$ is the same for all edges of all configurations of the underlying graph, allowing us to use the timescaled equation~\eqref{eq:edge-centric} between any two switching times. This remark applies in the context of discrete switching, but can be extended to a continuously-varying, weighted adjacency matrix, where $L(t)$ is no longer a piecewise constant matrix function.  
\end{rmk}

\subsection{Non-markovian random walks}
\label{sec:integro-diff}
As soon as the waiting time on the nodes, or the inter-activation times on the edges is no longer exponentially distributed, the Markov property is lost, and the differential equations~\eqref{eq:node-centric} and~\eqref{eq:edge-centric} are replaced by generalised integro-differential versions. Such generalisations have been developed from a node-centric perspective for instance in~\cite{angstmann2013pattern}, and for the edge-centric approach in~\cite{hoffmann2012generalized}, ending up in essentially the same form of equation, the only difference being the underlying mechanism regulating the residence times on the nodes. This lets us choose between the two approaches.
We will mainly follow  \cite{hoffmann2012generalized}, in which the generalised master equation valid for arbitrary distributions for $X_d$ in walk~(Mod. 2) is derived. This master equation will be an important ingredient in the next two sections. Therefore, it is worth introducing the equation, starting with some preliminary notations. 

The building quantity in a model is the residence time on a node $j$, namely the duration between the arrival-time on the node,  and a jump to any other node. This duration is a random variable $X^{(j)}$ with PDF $T_{\bullet j}(t)$.  This residence time density (also known as transition density)    satisfies the normalisation condition
\begin{equation}\label{eq:normalizationT}
\int_0^\infty  T_{\bullet j} (\tau ) d \tau  = 1, 
\end{equation}
meaning that a jump will eventually occur since the out-degree in the underlying graph is positive. The  diagonal matrix of the residence time densities is $D_{T} (t)$, so that the elements are given by $ \left[    D_{T} (t)   \right]_{ij} = T_{\bullet j} (t) \delta_{ij}$. 

The residence time density can be written as the sum $T_{\bullet j} (t) =  \sum_{i \in V_j} T_{ij}(t) $, where $T_{ij}(t)$ refers to a jump across edge $j \rightarrow i$. If $A_{ji} = 1$, the integral
\begin{equation}\label{eq:normalizationTij}
\int_0^\infty  T_{i j} (\tau ) d \tau  = \frac{1}{k_j}
\end{equation}
is thus the probability that at the time of the jump, the walker located on node $j$ selects node $i$ as destination. Let the matrix function   $\underbar T(t)$ have entries ${\underbar T}_{ij}(t) = T_{ij}(t)$. Finally, recall that the Laplace transform of the function $f(t)$ is the map   $s \mapsto \int_0^\infty f(t) e^{-st}dt$, and is written $\widehat{f}(s)$ or $\mathcal L \left\{ f(t) \right\}$. 

We are now in position to write the master equation. In the  Laplace domain it reads 
\begin{equation}\label{eq:Laplace}
\mathcal{L} \left\{\dot n(t)     \right\}     = \left(   \widehat {\underbar T} (s) \widehat D_{T}^{-1} (s) -I         \right) K(s) \widehat n(s), 
\end{equation}
where $\widehat K (s)   = \frac{s \widehat D_{T} (s) }{1- \widehat D_{T} (s) }$ is the memory kernel. The time-domain version of \eqref{eq:Laplace} is
\begin{equation} \label{eq:Time-Domain}
\dot n (t) = \left(               \underbar T(t) \ast  \mathcal L ^{-1} \left\{  \widehat D_{T}^{-1} (s)   \right\} - \delta (t)  \right) \ast K(t) \ast n(t). 
\end{equation}
This equation is  more involved than in the markovian case, and the analysis is better pushed further in the $s$-domain.
In particular, it will allow to obtain a compact expression for the steady state of the walk in section~\ref{sec:steady-state}. 

At this point, it is worth observing that equation~\eqref{eq:Time-Domain}   is generally non-Markovian in time, but it results in a   sequence of visited nodes that is indeed captured by a Markov chain. This means that it is  Markovian in trajectory. Therefore, it cannot be used without further assumption for model 4 that is generally Markovian neither in time, neither in trajectory.  Indeed,  we will see in section~\ref{sec:memory} that if there are cycles in the network, the next jump in model 4 along a cycle is conditional to  stochastic realisations in previous steps, hence affecting the choice of the next destination node.

The  work ahead  is now to compute the residence time  densities $T_{\bullet j}(t)$, from which the average time spent a node for a given model follows directly, as is shown in the next section.

\section{Mean residence times}
\label{sec:residence times}
The mathematical expectation the  residence time $X^{(j)}$ on node $j$ is given by
\begin{equation}
\label{eq:mrt}
E \big(X^{(j)}\big) = \int_0^\infty t T_{\bullet j } (t) dt, 
\end{equation}
and will also be referred to by $\langle T_{\bullet j} \rangle $. It is naturally called the mean residence time (or mean resting time) and is relevant in many scenarios, as it will for instance directly determine the relaxation time on tree-like structures\footnote{Our choice of studying the resting time is certainly restrictive, and quantities typically associated to random walks, such as the mean first-passage time or dispersal distribution, could be natural next steps. }.     We compute and interpret this quantity starting from the agent- and edge-level rules of the  different models, to obtain a macroscopic interpretation. Our analysis is restricted to exponential densities for $X_w$, $X_u$ and $X_d$, since this will allow to shed light on the effect of having up to three timescales, and not on complications arising from  otherwise possibly fat-tailed distributions for  these three random variables. As a results, the two edge-centric models 2 and 3 generate statistically equivalent trajectories, and the analysis for (Mod. 2) holds for (Mod. 3). 

\subsection{Derivation of the mean residence times}
In this section we handle the models by increasing level of complexity.

\paragraph{Models 1, 2 and 3.} In the active node-centric walk one is allowed to write directly $\langle T_{\bullet j} \rangle_\mathrm{model \ 1} = \frac{1}{\mu}$. When it comes to the active edge-centric walk (Mod. 2),  when instantaneous activation times follow a Poisson process, we have
$$
T_{ij}(t) = D (t) \left[       \int_t^\infty D(\tau) d \tau    \right]^{k_j-1}  . 
$$
The interpretation is that the edge $j \rightarrow i$ must activate after a time $t$, whereas all competing edges must remain unavailable at least up to that point.  Performing the integration and multiplying by $k_j$ gives
$
T_{\bullet j} = k_j \lambda e^{-k_j \lambda t}, 
$
a result already found in~\cite{hoffmann2012generalized}. This is again an exponential distribution with rate $k_j \lambda$. It follows that 
$$
\langle T_{\bullet j} \rangle_\mathrm{model \ 2} = E(\mathbb{ X}_d^{(j)}),
$$
where $\mathbb{ X}_d^{(j)}$ was introduced by equation~\eqref{eq:min}.

\paragraph{Model 6.} The walker residing in node $j$ will jump along edge $j \rightarrow i$ at time $t$ if all competing edges are unavailable at least up until then - that is, their period of unavailability will last for at least $t-x$, where $x$ marks the time the walker is ready to jump. Moreover, edge $j \rightarrow i$ needs to activate exactly after the duration $t-x$.  With this in mind, and when all distributions are exponential, it was shown in~\cite{petit2018random} that : 
\begin{align}
T_{ij}(t) &= \int_0^t \psi_j(x) \left[  \int_{t-x}^\infty D(s) ds    \right]^{k_j-1} D(t-x) dx ,\nonumber \\
\intertext{and noting that $\left[  \int_{t-x}^\infty D(s) ds    \right]^{k_j-1}$ simplifies to $e^{-\lambda (k_j-1)(t-x)}$,}
&=\int_0^t \mu e^{-\mu  x} \lambda e^{-k_j \lambda (t-x)} dx \label{eq:convolutionD}\\
&=\mu \lambda e^{-k_j \lambda t} \int_0^t e^{(-\mu + k_j \lambda)x} dx. \label{eq:D:Tij}
\end{align}
Note that \eqref{eq:convolutionD} is merely the convolution between the waiting time of the walker and the minimum of $k_j$ independent down-times for the edges, reflecting the fact that the process results in  an addition of random variables. To proceed, we observe that the integral in \eqref{eq:D:Tij} depends on whether $\mu = k_j \lambda$ or $\mu \neq k_j \lambda$. In the former case, the integral is equal to $t$ and multiplying \eqref{eq:D:Tij} by $k_j$ yields
$
T_{\bullet j} = \mu k_j \lambda e^{-k_j \lambda t} t. 
$
Hence, the mean residence time is 
\begin{equation}
\langle  T_{\bullet j} \rangle_\mathrm{model \ 6} = \int_0^\infty  \mu t^2 k_j \lambda e^{-k_j \lambda t} dt . 
\end{equation}
Recalling that the $n$-th moment of an exponential distribution with rate $\lambda$ is $E(X^n) = n!/\lambda ^n$, we have $\langle  T_{\bullet j} \rangle = 2/ \mu = 1/\mu + 1/(k_j \lambda) $. We will show that we get the same expression also in the second case, i.e. when $\mu \neq k_j \lambda$. Indeed \eqref{eq:D:Tij}  becomes
\begin{equation}
\label{eq:D:pdf}
T_{\bullet j} = \frac{\lambda \mu}{k_j \lambda - \mu } \left(   e^{-\mu t}    - e^{-k_j \lambda  t }      \right).
\end{equation}
The mean residence time follows from~\eqref{eq:D:pdf} : 
\begin{align}
\langle T_{\bullet j}  \rangle_{\mathrm{model \ 6}}  &= \frac{\lambda \mu}{k_j \lambda - \mu }  \int_0^\infty  \left(   e^{-\mu \tau}    - e^{-k_j \lambda \tau }      \right) d\tau \nonumber \\ 
&= \frac{\lambda \mu}{k_j \lambda - \mu }   \left(   \frac 1 {\mu^2} - \frac{ 1}{(k_j \lambda)^2}  \right) \nonumber \\
&=\frac 1 \mu + \frac{1}{k_j \lambda}, \label{eq:D:residence}
\end{align}
or also, $ \langle T_{\bullet j} \rangle_{\mathrm{model \ 6}}   = E(X_w) + E (       \mathbb X^{(j)}_d    )$, justifying again to consider (Mod. 6) an additive model.  

\paragraph{Model 4.} We have mentioned at then end of section~\ref{sec:integro-diff} that this model is generally non-Markovian in time, and also not Markovian in trajectories, a fact that will be further discussed in section~\ref{sec:memory}. However, we can already mention that no  memory  will arise in the choice of the next destination node if there are no cycles in the network. Therefore, the following derivation assumes a directed acyclic graph (DAG), which restores Markovianity in the trajectories, and equation~\eqref{eq:Laplace} can be used. Let us now determine the resting time density. In model 4,  two possible scenarios face the walker ready to jump~: either an edge is available (probability $r$), or an extra wait period is needed before an outgoing edge turns available (probability $1-r$). We have $r = \frac{\lambda }{\lambda + \eta}$ and $1-r  = \frac{\eta }{\lambda + \eta}$. It was therefore shown in~\cite{petit2018random} that $T_{ij}(t)$ has two terms, such that the transition density from node $j$  reads
\begin{equation} \label{eq:B1:Tdotj}
T_{\bullet j}(t) =  \left[  1-(1-r)^{k_j}    \right]  \psi_j (t) + (1-r)^{k_j} k_j\lambda \mu e^{-k_j \lambda t} \int_0^t e^{(-\mu + k_j \lambda)x}dx. 
\end{equation}
The two terms reflect  a weighted combination of models 1 and 6. The weight $(1-r)^{k_j}$ is the probability that all outgoing edges are unavailable at a random time.  It follows that 
\begin{align}
\langle T_{\bullet j}  \rangle_{\mathrm{model} \ 4} &=   \langle T_{\bullet j}  \rangle_{\mathrm{model} \ 1} 
+  \langle T_{\bullet j}  \rangle_{\mathrm{model} \ 6} \nonumber \\
&= \left[  1-(1-r)^{k_j}    \right]  E(X_w) + (1-r)^{k_j}  \left(  E(X_w) + E(\mathbb{ X}_d^{(j)} ) \right)\nonumber \\
&= E(X_w) +  (1-r)^{k_j}   E(\mathbb{ X}_d^{(j)} )  \label{eq:B1:residence}. 
\end{align}
Under this form, we see the model is conditionally (depending on $r$) additive.

\paragraph{Model 5.} When the walker is ready to jump, the availability of network edges depends on the duration since the walker arrived on the node. That makes the analysis somewhat more involved. Assume the walker is ready after $s$ time units. Let $p^*(s)$ be the probability that an edge is in the same state it was at time $t = 0$, namely, unavailable. Let also $q^*(s) = 1-p^*(s)$ be the probability the edge is available for transport. These two quantities were computed in \cite{petit2018random}, by accounting for all possible on-off switches of the edge in the interval $[0,s]$. The resulting expression has a strikingly simple form when $U(t)$ and $D(t)$ have the same (exponential density) rate $\eta = \lambda$, our working hypothesis in what follows : 
\begin{align}
p^*(s) & = \frac{1}{2} (1+ e^{-2\lambda t}), \label{eq:pstar}\\
q^*(s) & = \frac{1}{2} (1- e^{-2\lambda t}). \label{eq:qstar}
\end{align}
If the walker is ready after a short time $s$, the edge will probably still be down, $p^*(0)  = 1$, while for large $s$, the state of the edge is up or down with equal probability, \mbox{$\lim_{s \rightarrow \infty } p^*(s) = \frac 1 2 $}. 

So now, we have an expression similar to \eqref{eq:B1:Tdotj} except that $r$ and $1-r$ are  essentially replaced by the time-dependent $q^*$ and $p^*$. Let us begin by first writing an expression for  $T_{ij}(t)$ : 
\begin{align}\label{eq:C:Tij}
T_{ij}(t) &= \frac{1}{k_j} \left[ 1-p^*(t)^{k_j}      \right]\psi_j(t) +  \int_0^t\psi_j(x)     \left[  p^*(x) \int_{t-x} D(s) ds\right]^{k_j-1} p^*(x) D(t-x) dx \nonumber \\
&= \frac{1}{k_j} \left[ 1-p^*(t)^{k_j}      \right]\psi_j(t) + \lambda \mu e^{-k_j \lambda t} \int_0^t p^*(x)^{k_j} e^{(-\mu + k_j \lambda ) x}dx \nonumber \\
&= \frac{1}{k_j} \left[ 1-p^*(t)^{k_j}      \right]\psi_j(t) + \frac{\lambda \mu }{2^{k_j}}e^{-k_j \lambda t} \int_0^t \left(1+e^{-2\lambda x}\right)^{k_j}e^{(-\mu + k_j \lambda ) x}dx, 
\intertext{and using Newton's binomial formula in both terms,   }
&=  \frac{1}{k_j}     \left[ 1-       \frac{1}{2^{k_j} }   \sum_{m=0}^{k_j}        \binom{k_j}{m} e^{-2m\lambda t}        \right]\psi_j(t) +  \frac{\lambda \mu }{2^{k_j}}e^{-k_j \lambda t}  \sum_{m = 0}^{k_j} \binom{k_j}{m} \frac{1}{\beta_m} \left(  e^{\beta_m t} -1   \right)
\end{align} 
where we have set  $\beta_m = -\mu + k_j \lambda -2m \lambda $. The resting time density therefore reads : 
\begin{multline} \label{eq:C:Tdotj}
T_{\bullet j}  (t) = \mu e{-\mu t} - \frac{\mu }{2^{k_j}}   \sum_{m=0}^{k_j}        \binom{k_j}{m} e^{-(\mu+2m\lambda) t}   \\
+ \frac{\mu}{2^{k_j}} k_j \lambda   \sum_{m=0}^{k_j}        \binom{k_j}{m} \frac{1}{\beta_m} e^{-(\mu +2 m\lambda ) t}
- \frac{\mu}{2^{k_j}} k_j \lambda   \sum_{m=0}^{k_j}  e^{-k_j \lambda t}      \binom{k_j}{m} \frac{1}{\beta_m} . 
\end{multline}
The mean resting time follows as 
\begin{multline}
\langle T_{\bullet j}  \rangle_{\mathrm{model} \ 5}  = E(X_w)  - \frac{\mu }{2^{k_j}} \sum_{m = 0}^{k_j}  \binom{k_j}{m} \frac{1}{(\mu + 2 m \lambda)^{2}}   \\
+ \frac{\mu}{2^{k_j}    }    k_j \lambda   \sum_{m = 0}^{k_j}  \binom{k_j}{m} \frac{1}{\beta_m}  \frac{1}{(\mu + 2 m \lambda)^{2}}  
- \frac{\mu}{2^{k_j}    }   \frac{1}{k_j \lambda }    \sum_{m = 0}^{k_j}  \binom{k_j}{m} \frac{1}{\beta_m}. 
\end{multline}
Regrouping the terms, we get 
\begin{align}
\langle T_{\bullet j}  \rangle_{\mathrm{model} \ 5}  &= E(X_w)  + \frac{\mu}{2^{k_j}}    \sum_{m = 0}^{k_j}  \binom{k_j}{m}  \left(         
\frac{1}{(\mu + 2m \lambda)^2}    \left[   \frac{k_j \lambda}{\beta_m} -1   \right] - \frac{1}{\beta_m} \frac{1}{k_j \lambda} 
\right)     \nonumber \\
&= E(X_w)  + \frac{\mu}{2^{k_j}}    \sum_{m = 0}^{k_j}  \binom{k_j}{m} \frac{1}{\beta_m} \left(     \frac{1}{\mu + 2 m \lambda }   - \frac{1}{k_j \lambda}    \right) \nonumber \\
&= E(X_w)  + \frac{\mu}{2^{k_j}     }   \sum_{m = 0}^{k_j}  \binom{k_j}{m} \frac{1}{\mu +2 m \lambda}  E(\mathbb{ X}_d^{(j)} )  . 
\label{eq:C:restingtime}
\end{align}

\subsection{Discussion}

All models have a mean residence time that can be cast under the form
\begin{equation}
\langle T_{\bullet j}  \rangle_{\mathrm{model}} = a_{\mathrm{model}} E(X_w)    + b_{\mathrm{model}}(k_j,\mu, \lambda) E(\mathbb{ X}_d^{(j)} ), 
\end{equation}
where $a_{\mathrm{model}} = 1$ for all models but (Mod. 2) for which it is 0, and $b_{\mathrm{model}}(k_j, \mu, \lambda)$ accounts for the probability that all outgoing edges are unavailable when the walker is ready to jump. Summing up the results of this section, we have
\begin{align}
&b_{\mathrm{model \ 1}} (k_j,\mu, \lambda) =     0     \\
&b_{\mathrm{model \ 4}} (k_j,\mu, \lambda) =      (1-r)^{k_j}   \\
&b_{\mathrm{model \ 5}} (k_j,\mu, \lambda) =        \frac{1}{2^{k_j}} \mu \sum_{m =0}^{k_j} \binom{k_j}{m}  \frac{1}{\mu + 2m \lambda} \label{eq:bmodc}\\
&b_{\mathrm{model \ 6}} (k_j,\mu, \lambda) =   1. 
\end{align}
Recall that \eqref{eq:bmodc} was derived under the assumption that $\eta = \lambda$, for which $r = \frac{1}{2}$ and thus $b_{\mathrm{model \ 4}}(k_j,\mu, \lambda) = \frac{1}{2^{k_j}}$. Using standard algebra, it is straightforward to check that 
\begin{equation}
0 = b_{\mathrm{model \ 1}}  < b_{\mathrm{model \ 4}}  < b_{\mathrm{model \ 5}} < b_{\mathrm{model \ 6}} = 1, 
\end{equation}
for all $k_j \in \mathbb{ N}_0$ and all positive reals $\mu$ and $\lambda$. The smaller this coefficient, the larger the expected number of jumps along the trajectories of the walk, all other parameters being chosen equal.

We want to compare the three models with nonzero $b_{\mathrm{model }}$, since these are the ones where there is a dynamical walker-network interaction. To this end, let us define the ratios of mean residence times 
\begin{align}
R_1 &: =   \frac{\langle T_{\bullet j}  \rangle_{\mathrm{model\ 4}}}{\langle T_{\bullet j}  \rangle_{\mathrm{model \ 6}}}  \ ,  \label{eq:R1def}\\
R_2 & :=  \frac{\langle T_{\bullet j}  \rangle_{\mathrm{model\ 5}}}{\langle T_{\bullet j}  \rangle_{\mathrm{model \ 6}}}.  \label{eq:R2def}
\end{align}
These quantities depend only on the degree $k_j$, and on a new variable $\xi := \frac{\lambda}{\mu}$. Indeed, we can write
\begin{align}
R_1(k_j, \xi) & =    \frac{ k_j 2^{k_j}\xi + 1}{k_j 2^{k_j}\xi + 2^{k_j}} \ ,  \label{eq:R1}\\
R_2(k_j, \xi) & =  \frac{ k_j 2^{k_j}\xi +\sum_{m = 0}^{k_j}  \binom{k_j}{m}   \frac{ 1}{1+2m \xi}   }{k_j 2^{k_j}\xi + 2^{k_j}}.  \label{eq:R2}
\end{align}

\begin{figure}
	\centering
	\hspace*{-.5em}\includegraphics[width=\textwidth]{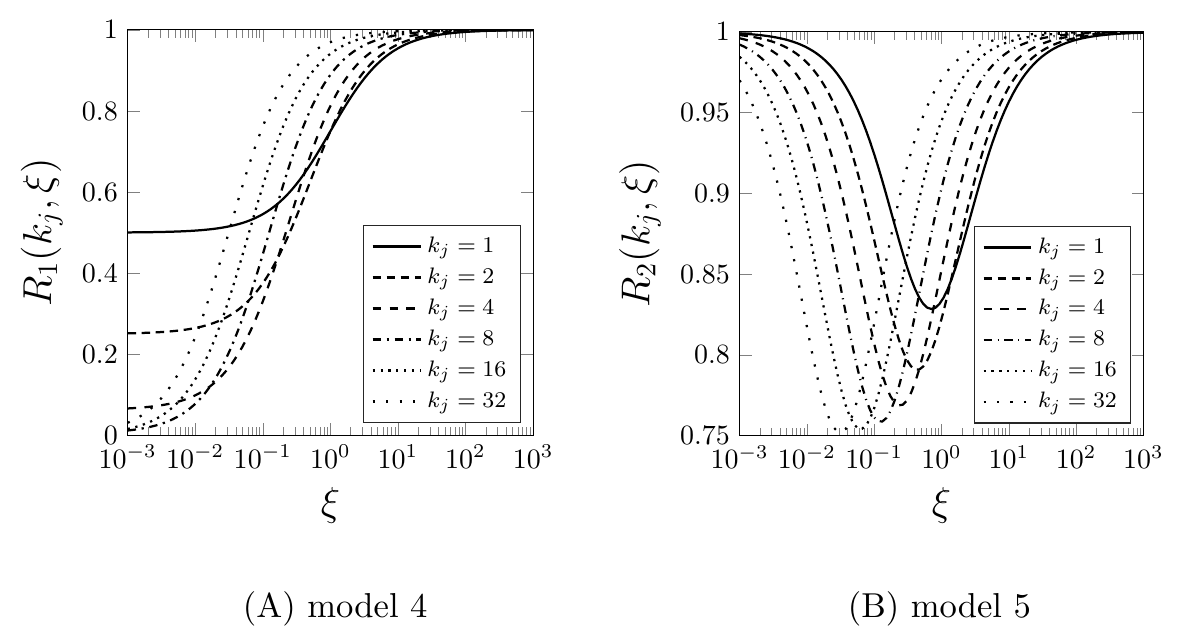}
	\caption{\textbf{Mean residence times with respect to model 6.}  Going from model (1) to (4) to (5) to (6), the mean residence time is increasing - all dynamical and topological parameters being equally chosen, and diffusion on, say, a tree topology would be slower. Ratios~$R_1$ and~$R_2$ of equations~\eqref{eq:R1} and~\eqref{eq:R2} allow to compare the former three models with  two timescales (corresponding to $\mu$ and $\lambda = \eta$),  for which there is walker-edges   interaction. It measures  the reduction in  mean residence times of walks (Mod. 4) and (Mod. 5) with respect to the additive model 6. Observe that only the degree and  $\xi = \frac  \lambda \mu$ determine these comparisons. That $R_i(k_j, \xi \rightarrow \infty) = 1$ for $i = 1,2$ and $R_2(k_j, 0) = 1$ is explained by the asymptotic behaviour described in  figure~\ref{fig:arrows}. Also observe the different ranges of  values for $R_1$ and $R_2$. The smallest value of $R_2$  is above  $\frac{3}{4}$, when $k_j \rightarrow \infty$, whereas $R_1$ approaches zero.   }
	\label{fig:residencetime}
\end{figure}

\begin{figure}[htb!]
	\centering
	\includegraphics[width=.8\textwidth]{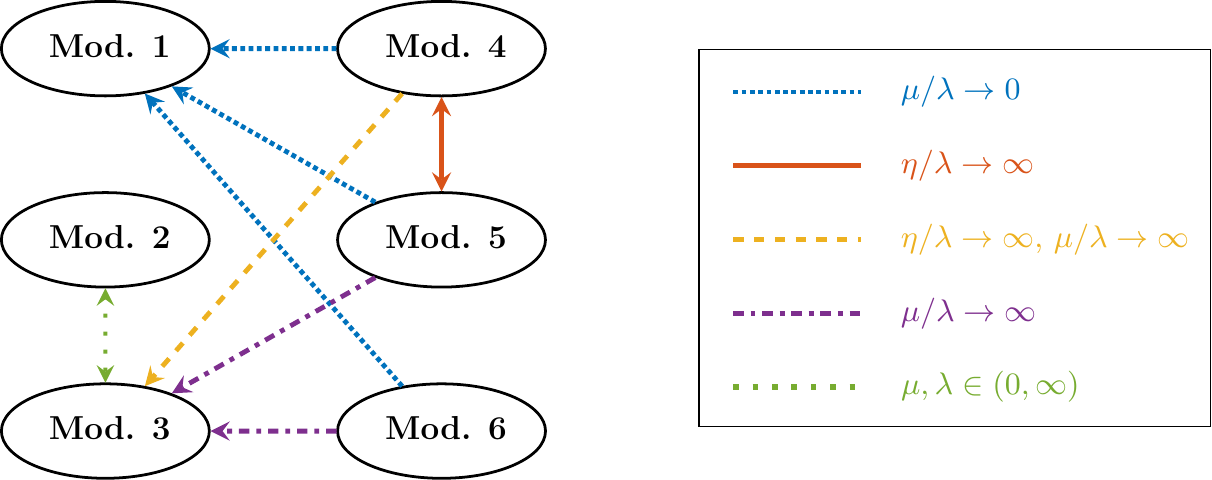}
	\caption{\textbf{Relationship between models with all-exponential distributions} 
		The  graph represents the relationships between all models when all durations are exponentially distributed. Here, $\mu$, $\eta$ and $\lambda$ represent the exponential rates of the distributions, such  that $\langle X_w \rangle = \frac{1}{\mu}$ for the walker,  $\langle X_u \rangle = \frac{1}{\eta}$ for the up times, and  $\langle X_d\rangle = \frac{1}{\lambda}$ for the down times. The arrows represent  pointwise convergence for all times, of   the resting time PDF's $T_{ij}(t)$ in function of the dynamical parameters. As indicated in the main text,  convergence regarding  model~5 was established when $\eta = \lambda$.  Also note that (Mod. 2) and (Mod. 3) can be merged in the all-exponential case since they produce statistically indistinguishable  trajectories.}
	\label{fig:arrows}
\end{figure}

The above expressions are plotted in figure~\ref{fig:residencetime} for various values of the degree. The reduction in mean resting time for model 4 is very pronounced for small $\xi$, especially for the large degrees for which the relatively slow network timescales have less effect. With model~5 however,   the reduction factor never goes below $\frac{3}{4}$.  In terms of convergence of the models, we observe that for all degrees, $R_1 \approx 1$ for large $\xi$, and $R_2 \approx 1$ for both large and small $\xi$. This behaviour  for large $\xi$ is  a direct consequence of the convergence of the resting time PDF's of models 4, 5 and 6  to that of model~1 when $\xi^{-1} = \frac{\mu}{\lambda}  \rightarrow 0$. This is represented by the three blue dotted arrows on figure~\ref{fig:arrows}. On the other hand, the value of $R_2$ for small $\xi$ results from the convergence indicated by the two purple dash-dotted arrows of the figure.  The other arrows further indicate the convergence between the PDF's $T_{ij} (t)$ of the different models in  asymptotic regimes of the dynamical parameters $\mu, \eta, \lambda$. These results can be verified  by direct computation from the  expressions for the densities we have obtained in this section.  	Obviously,   convergence of the densities implies convergence  for the expectations. For instance, consider the blue arrow from (Mod. 5) to (Mod. 1). In terms of mean resting time, we have that when $\lambda \rightarrow \infty, \mu \in \mathbb{ R}$, then $ \langle T_{\bullet j}  \rangle_{\mathrm{model} \ 5}  \rightarrow E(X_w)$. If we now let $\mu \rightarrow \infty, \lambda \in \mathbb{ R}$, then the mean tends to $   \frac{1}{2^{k_j} k_j \lambda}  \sum_{m = 0}^{k_j}  \binom{k_j}{m}  = \frac{1}{k_j \lambda} $, that is, $E(\mathbb{ X}_d^{(j)} ) $ (purple arrow from (Mod. 5) to (Mod. 3)). In both cases, this is the expected outcome.

\section{Steady state}
\label{sec:steady-state}
\begin{figure}[h!]
	\centering
	\hspace*{-.2cm}\includegraphics[width=\textwidth]{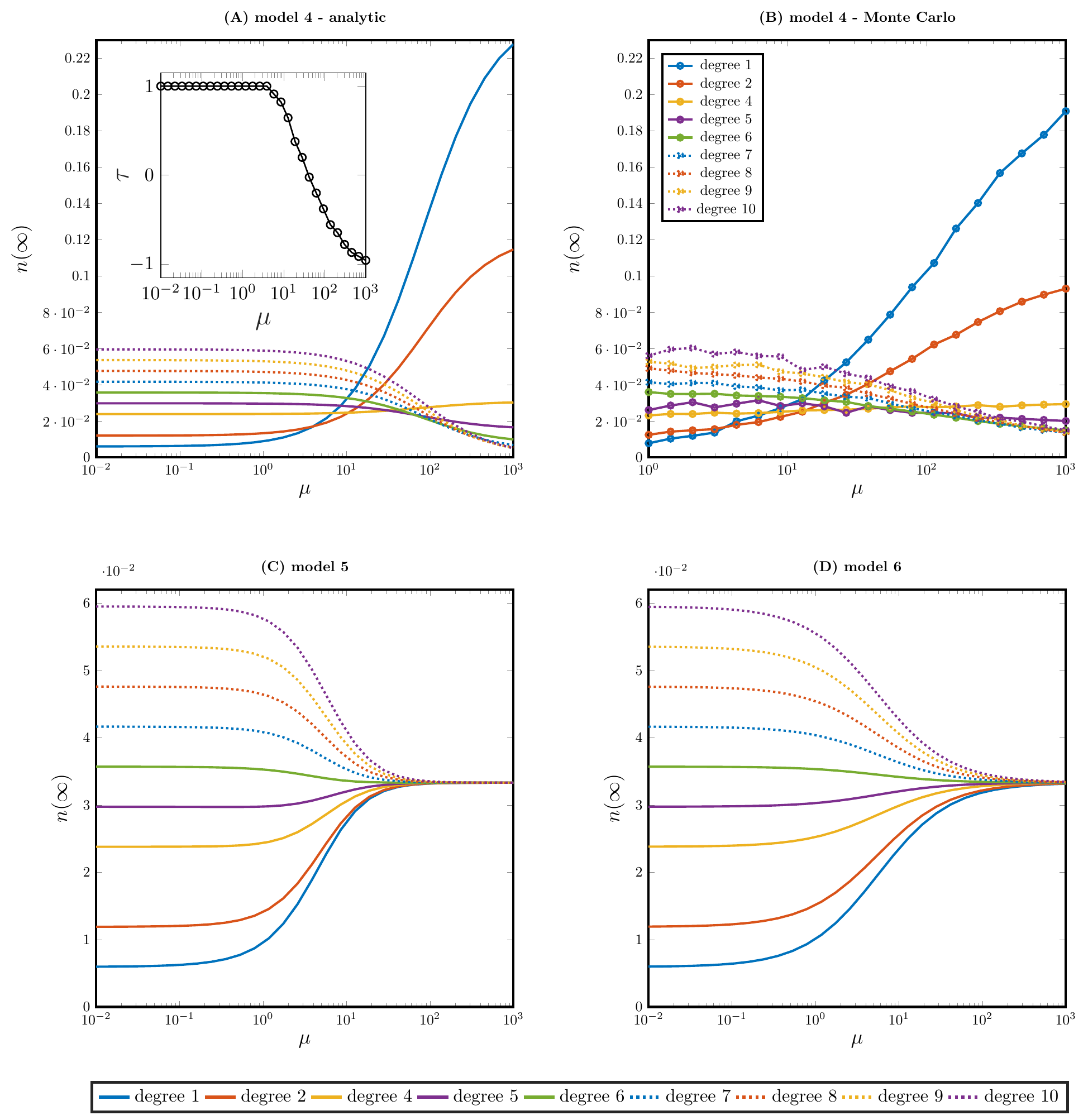}
	\caption{\textbf{Steady states as a ranking application.}  
		Models 5 and  6 on the bottom panels (C) and (D) behave in the limit for the fast walker ($\mu \rightarrow \infty $)  as a Markovian passive edge-centric random walk, where the steady state is uniform across the symmetric network. In the small-$\mu$ limit, the steady state probability  is proportional to the degree, like in the Markovian active node-centric walk. In model 4 of panel (A), this behaviour does not hold true. Indeed,  when $\mu \rightarrow \infty$, the steady-state residence probability  is high when degree is low. This inversion with respect to the degree-based ranking is captured by the inset of panel (A). On this inset, the evolution of $\tau$, which denotes Kendall's Tau coefficient, is represented in function of $\mu$. When Kendall's Tau is one, the rankings of panels (A) and (C) or (D) are the same; when it is -1, the two rankings are reversed. There is no link between the two when it is 0. Panel (B) displays the result of Monte-Carlo simulation of $10^4$ trajectories of a walker subject to the motion rules of model 4, in the range of values of $\mu$ corresponding to the reversed ranking. These simulations do account for the presence of cycles, and confirm quantitatively the  ranking predicted by the analytical formula.  We have selected $\eta = 1  = \lambda $ for all plots. The strongly connected symmetric network is Erd\"os-R\'enyi with 30 nodes and connection probability $\frac{1}{5}$. }
	\label{fig:ss}
\end{figure}

The steady state of the walk, $  n( \infty) : = \lim_{t \rightarrow \infty} n(t)$, is a  useful quantity for instance in  ranking applications. This motivates the content of this section. The steady state  can be obtained based on the mean residence time of the preceding section. Let $ D_{\langle T  \rangle}  $ be the diagonal matrix containing the mean waiting times on the nodes, such that $ \left[D_{\langle T  \rangle}\right]_{ij} = \langle T_{\bullet j} \rangle  \delta_{ij}$.  In~\cite{hoffmann2012generalized}, a small-$s$ analysis of the generalised master equation~\eqref{eq:Laplace} showed the steady-state of the walk to be
\begin{equation}\label{eq:stationarystate}
n( \infty) : = \lim_{t \rightarrow \infty} n(t) \propto D_{\langle T \rangle} v, 
\end{equation}
where $v$ is the eigenvector associated to the unit eigenvalue of the effective transition matrix $\mathbb{T}$. This matrix has elements 
$$\mathbb{T}_{ij} = \int_0^\infty T_{i \leftarrow j} (t) dt =  \frac{1}{k_j}A_{ji}. $$ 
We can straightforwardly check that $v_j = k_j$ satisfies $\mathbb (T v)_i= \sum_{\ell= 1}^N A_{\ell i} = k_i$, where the last equality assumes that the network is balanced, namely, the in-degree of node $i$ is equal to its out-degree $k_i$. In other words, when the underlying network is balanced, the steady state is~\cite{hoffmann2012generalized} 
\begin{equation}\label{eq:steadystate-symmetric}
n_j( \infty ) =  \alpha D_{\langle T  \rangle} k_j, 
\end{equation} 
where $\alpha $ is the normalisation factor. As mentioned before, one interesting application of random walks is their use to rank nodes of a network according to the probability to find a walker on each node, an information directly accessible from the  steady state. We are thus interested in understanding how the   asymptotic states depend on the   modeling scheme and hence how the ranking process changes accordingly. We compute the steady state for each model in the sequel, and report the results graphically on figure~\ref{fig:ss}. 

\paragraph{Models 1, 2 and 3.} 
For the sake of completeness and for later comparisons,  we recall some standard results.  For the active node-centric model 1, $\langle T_{\bullet j} \rangle  = \frac 1 \mu $ and the steady-state is proportional to degree, 
$$n_j^{(Mod.\ 1)} (\infty ) = \frac{k_j}{\sum_{\ell = 1}^N k_l}.$$
The active edge-centric model~2 yields 
$$n_j^{(Mod. \ 2)} (\infty ) = \frac{1}{N}.$$
As already pointed out, the same expression if valid for the passive edge-centric walk (Mod. 3) when the down-time distribution for network edges are exponential. One can easily verify that the right-hand side of $\dot n = n L^{rw}$ vanishes for $n = n_j^{(Mod. \ 1)} (\infty )$, while the same holds true with $\dot n = n L$ for $n = n_j^{(Mod. \ 2)} (\infty )$.

\paragraph{Model 6}
We have 
$
n_j^{(Mod. \ 6)} (\infty ) = \alpha^{(Mod. \ 6)} \left( \frac{1}{\mu} + \frac{1}{k_j \lambda}        \right) k_j = \frac{ \alpha^{(Mod. \ 6)} }{\lambda \mu }   ( \mu   + k_j \lambda) ,
$
and after normalisation we get 
\begin{equation}
n_j^{(Mod. \ 6)} (\infty )  = \frac{k_j \lambda + \mu }{\sum_{\ell = 1}^N k_\ell \lambda + \mu }
\end{equation}
or under a different form, after division by $k_j \lambda \mu $, 
\begin{equation}
n_j^{(Mod. \ 6)} (\infty )  = \frac{\frac{1}{\mu } + \frac{1}{k_j \lambda}}{\sum_{\ell = 1}^N \frac{k_\ell}{k_j}\frac{1}{\mu} + \frac{1}{k_j \lambda}} %
=\frac{E(X_w) + E(\mathbb{X}_d^{(j)}  )  }{\sum_{\ell = 1}^N \frac{k_\ell}{k_j}E(X_w)  +  E(\mathbb{X}_d^{(j)}  )   } 
\end{equation}
We recover the expressions of  the active node-centric (Mod. 1) and  edge- centric (Mod. 4)  walks in the respective limits $\lambda \rightarrow \infty$ and $\mu \rightarrow \infty$.

\paragraph{Model 4}
This model  necessitates a preliminary observation concerning our method. We use the transition density derived in the preceding section,  which results to be only an approximation when the graph has cycles (see section~\ref{sec:memory}). We also rely on the steady state formula~\eqref{eq:steadystate-symmetric} obtained for balanced networks. Hence, the steady state   we will obtain with equation~\eqref{eq:B1:steadystate} is an approximation if the  balanced network has cycles, for instance when the network is symmetric. On figure~\ref{fig:ss}, we have indeed selected a symmetric network. The figure shows that the steady state computed with the theoretical formula is in good agreement  with Monte-Carlo simulations. The formula proved mostly accurate throughout our numerical investigation,   certainly in terms of ranking of the nodes. The network topologies and range of values for the dynamical parameters for which this observation holds, are further discussed in section~\ref{sec:memory}.

Let us  now proceed with the analysis. 
We have 
$$n_j^{(Mod. \ 4)} (\infty)= \alpha^{(Mod. \ 4)} \left(         \frac{k_j}{\mu} + (1-r)^{k_j} \frac{1}{\lambda}         \right)$$ 
and through normalisation we obtain
\begin{equation}\label{eq:B1:steadystate}
n_j^{(Mod. \ 4)} (\infty)=      \frac{  \frac{k_j}{\mu} + (1-r)^{k_j} \frac{1}{\lambda}   }{%
	\sum_{\ell = 1}^{N}   \left[    \frac{k_\ell}{\mu} + \frac{(1-r)^{k_\ell} }{\lambda} \right]
}.
\end{equation}
As expected, $n_j^{(Mod. \ 4)}(\infty)$ tends to  $n_j^{(Mod. \ 1)}(\infty)$ when $\lambda \rightarrow \infty$. But more importantly,  in the limit  of a very fast walker we have
$$\lim_{\mu \rightarrow \infty} n_j^{(Mod. \ 4)}(\infty) =  \frac{(1-r)^{k_j}}{\sum_{\ell = 1}^{N} (1-r)^{k_\ell}}. $$
It results  that smaller residence probabilities are associated with nodes with larger degree. At variance, for typically large walker waiting times, larger degree means larger probability. Indeed, observe that 
$$\lim_{\mu \rightarrow 0}  n_j^{(Mod. \ 4)} (\infty) = \frac{k_j }{\sum_{\ell = 1}^N k_\ell} . $$ 
It was reported before that fat-tailed residence times on a portion of the  nodes of a network could lead to accumulation on these nodes in spite of their low degree~\cite{fedotov2017anomalous}. In our case, the renewal process ruling the jump times arises from interaction between walker and network, without explicitly reverting to long-tailed distributions of the residence time on certain nodes. 

\paragraph{Model 5} 
Resulting directly from the transition density given by  \eqref{eq:C:restingtime}, we have 
\begin{equation}\label{eq:C:steadystate}
n^{(Mod. \ 5)} (\infty) = \alpha^{(Mod. \ 5)} \left(           \frac{1}{\mu} + \frac{\mu}{2^{k_j} \lambda}     \sum_{m = 0}^{k_j} \binom{k_j}{m}   \frac{1}{\mu + 2 m \lambda}      \right)
\end{equation}
with 
\begin{equation}
\alpha^{(Mod. \ 5)} = \left[          \frac N \mu    + \frac \mu \lambda  \sum_{\ell = 1}^N   
\left (  \frac{1}{2^{k_\ell}}   
\sum_{m = 0}^{k_\ell}           \binom{k_\ell}{m}   \frac{1}{\mu + 2 m \lambda}       \right )        \right]^{-1} . 
\end{equation}

\section{Memory through walker-network interaction}
\label{sec:memory}

In general,  the passive model  2  or the  passive-at-edge-level model 4  are not  tractable analytically, in the sense that without further assumptions each jump (time and destination) depends on the full trajectory of the walker. Model 2 becomes tractable assuming exponential distributions for the walker and edges, which allowed us to find the residence time density in section~\ref{sec:residence times}.  But this assumption is not enough in model 4 (unless we make the extra assumption of a directed acyclic network). 

Let us elaborate on this. We can make the choice that edges are statistically independent, in the sense that no correlation exists between the states (available or not) of the edges.  In this way, in all models but (Mod. 4), we  avoid preferred diffusion paths arising from correlations, as captured by the concepts of betweenness preference \cite{scholtes2014causality} or memory network \cite{rosvall2014memory}.
Model 4 is different.  Indeed, if the walker choose to  use an outgoing edge in the past, this gives information on the state of the same outgoing edge later in time.  This holds true even with exponential distributions for the up- and down-times. More precisely, let us consider two cases :  
\begin{itemize}
	\item A jump across an edge some time in the past (meaning it was available back then) increases the probability that the same edge is again available to the walker who has returned on the node after a cycle, and is ready for another jump.  This memory that impacts the state of the edge was described in details in~\cite{petit2018random}, and is captured by the duration-dependent probability $p^*$ in equation~\eqref{eq:pstar}. 
	\item Conversely, if an outgoing edge was not selected and lost the competition to another edge, this raises the probability that this edge was actually not available at the time of the jump (and hence explaining why it lost the competition). Therefore, when the walker returns on the node after a cycle,  with relatively more probability the same edge is not available when the walker wants to jump. Again, an expression similar to that of $p^*$ was obtained in ~\cite{petit2018random} to quantify this phenomenon. 
\end{itemize}
Combining these two cases shows that by the presence of cycles, preferred diffusion paths can emerge.

The impact of cycles may depend on many factors (rates, topology of the network, presence of communities, initial condition of the walk), and it is difficult to analyse and predict. For instance, from numerical experiments (data not reported here) we found a dependence on the number of cycles in the network, or the mean degree or also the number of nodes. It is safe to say however, that the effect decreases for long cycles, because the walker expectedly takes longer to complete the cycle. In the same line of thought, the effect is less pronounced if the rate $\mu$ of the walker is relatively small, because again, more time will elapse between the jumps, thereby reducing the memory. 
In general, larger networks have more possible diffusion paths, and tend to be less sensitive to the effect. 

Let us consider figure~\ref{fig:ss} again.  The difference between the analytical curves of panel (A) and the outcome of Monte Carlo simulations on  panel (B) is a consequence of the presence of the cycles.  It is not due to the variance in the simulation of the stochastic process, and was moreover observed for various topologies.   As anticipated, the effect  in this example is more pronounced with larger values of $\mu$. This particular example   demonstrates the existence of a memory effect, that is, a  bias in the trajectories and in terms of  rate of jumps,  due to  the interaction of a memoryless walker, evolving on a network governed by memoryless distributions. Markovianity both in time and in  trajectories is lost due to the microscopic rules of the model, and not due to the choice of the distribution or any explicit bias by the walker.  We point to our previous work in~\cite{petit2018random} for a mathematical framework further enabling an analytic approach. 

\section{Conclusion}
\label{sec:conclusion}

Many types of random walks have been defined on static networks, from correlated walks \cite{renshaw1981correlated} to other variants of elephant random walks or random walks with memory \cite{rosvall2014memory}. The
main purpose of this paper was to show that even the simplest model of random walk, where the walker does not have a memory and is unbiased, may generate complex trajectories when defined on temporal networks.
As we have discussed,  there exist different ways to inter-connect the dynamics of the walker and of the network, and this interplay  may  break the Markovianity of the system  (in time), even in purely active models or passive models without cycles. Note that there is no need for a long-tailed walker waiting time on (a subset of) the node(s), such as in \cite{fedotov2017anomalous}, to observe a dramatic departure from the  steady-states of the classical random walk models. We  also showed that the mean residence time may be impacted, resulting in a slowed-down diffusion on tree-like structures. The Markovianity of the trajectories of the walkers may also be broken when the underlying graph is not acyclic, as certain jumps are preferred based on the past trajectory of a walker, even if edges are statistically independent. 

Overall, our work shows the importance of the different timescales associated to random walks on temporal networks, and  unveils the importance of the duration of contacts on diffusion. We have also enriched the taxonomy of  random walk processes~\cite{masuda2017random}, adding to the known ``active'' walks, appropriate for human or animal trajectories and ``passive" walks, typically used for virus/information spreading  on temporal network, new combinations of active and passive processes that are relevant in situations when an active agent is constrained the dynamical properties of the underlying network. Typical examples would include the mobility of individuals on public transportation networks. Despite its richness, our model neglects certain aspects of real-life networks that could lead to interesting research directions. In particular, the implicit assumption that the network can be described as a stationary process calls for generalisations including  circadian rhythms~\cite{jo2012circadian,kobayashi2016tideh}. Another interesting generalisation would be to open up the modelling framework to situations when the number of diffusing entities is not conserved, and evolves in time, as in epidemic spreading on contact networks, where an additional temporal process is associated to the distribution of the recovery time of infected nodes~\cite{keeling1997disease}.

\section*{Competing interests}
  The authors declare that they have no competing interests.

\section*{Author's contributions}
J.P., R.L. and T.C.  conceived the project and wrote the manuscript. J.P. derived the analytical results and performed the numerical simulations. 

\bibliography{ANS_refs.bib}      

\end{document}